\documentclass[article,twocolumn,showpacs,superscriptaddress,10pt]{revtex4-1} 
\usepackage{siunitx,graphicx,overpic,epstopdf,amsmath,amssymb,bm,color}
\usepackage{tikz,tkz-tab}
\usepackage{subcaption} 
\usepackage{hyperref}
\usepackage{subcaption} 
\hypersetup{hidelinks}

\begin{document}

\title{Fast and High Quality Light Field Acquisition using Defocus Modulation}

\author{Haichao Wang}
\affiliation{Shanghai Institute of Optics and Fine Mechanics, Chinese Academy of Sciences,	Shanghai 201800, China}
\affiliation{University of the Chinese Academy of Sciences}

\author{Ni Chen}\email{Corresponding author: nichen@siom.ac.cn}
\affiliation{Shanghai Institute of Optics and Fine Mechanics, Chinese Academy of Sciences,	Shanghai 201800, China}

\author{Jingdan Liu}
\affiliation{School of Optoelectronics, Beijing Institute of Technology, Beijing 100081, China}

\author{Guohai Situ}\email{Corresponding author: ghsitu@siom.ac.cn}
\affiliation{Shanghai Institute of Optics and Fine Mechanics, Chinese Academy of Sciences,	Shanghai 201800, China}

\begin{abstract}
Light field reconstruction from images captured by focal plane sweeping, such as light field moment imaging~(LFMI) and light field reconstruction with back projection~(LFBP), can achieve high lateral resolution comparable to the modern camera sensor. This is impossible for the conventional lens array based light field capture systems. However, capturing a series of focal plane sweeping images along the optical axis is time consuming and requires fine alignment. Besides, different focal plane based light field reconstruction techniques require images with different characteristics. To solve these problems, we present an efficient approach for fast light field acquisition with precise focal plane sweeping capture by defocus modulation rather than axial movement. Because of the controllable point spread function, we can capture images for light field reconstruction with both LFMI and LFBP. 
\end{abstract}


\maketitle 

\section{Introduction}
Generally, a conventional imaging systems record intensity-only images, while the depth information of the three-dimensional~(3D) scene is lost. However, the depth information can be extracted from the light field, which records not only the intensity but also the propagation directions of the light rays~\cite{levoy2006light}. Generally, the light field can be captured by either a lens array with a standard camera~\cite{ng2005light, levoy2006light} or a camera array~\cite{wilburn2005high,lin2015camera}. From the view of geometric optics, those methods simultaneously record the two-dimensional~(2D) spatial and angular information of the light rays, thus allowing perspective view image generation, refocusing of the scene, and free-glass 3D display~\cite{hong2011three, ng2005light,park2014recent}. However, lens array based light field capture~\cite{levoy2006light, wilburn2005high,lin2015camera, hong2011three, ng2005light, park2014recent,levoy2009recording,prevedel2014simultaneous} has to make an intrinsic trade-off between the the spatial and the angular resolution. This is because when the size of the lenslet is large, one of the captured elemental image will have a large spatial resolution, therefore the covered quantity of lenslet that the light rays from the object scene will be small, which leads to less number of elemental images, i.e., low angular resolution. Although there exists some techniques to improve the resolution~\cite{chen2011resolution,Chen_2010_OE}, the trade-off induced by the lens array can not be break through.

Coded masks inserted into a camera has also been invented to obtain a higher resolution light field. However, it sacrifices the light transmission because of the attenuation induced by the masks~\cite{Veeraraghavan_2007_ACM,Marwah_2013_ACM}. Recently, it has been reported that the light field can also be obtained from a series of focal plane sweeping captured images with a conventional digital camera~\cite{orth2013light,park2014light,Mousnier_2015_ARXIV}. These techniques can obtain a higher resolution light field. In these cases, the light field is calculated from several photographic images captured at different focal planes, the images are not segmented by the sub lenslet of the lens array, hence reach a higher angular and spatial image resolution comparable to a conventional camera sensor. As these methods do not require any special equipments like lens array or code masks, they are easy to be implemented. However, they require a large stack of defocused images to research an accurate light field reconstruction~\cite{park2014light,liu2015light,Chen_2017_AO,Yin_2016_AO}, in which the capture process is time consuming and requires fine alignment. In this paper, we propose an efficient technique for fast, precisely focal plane sweeping capture with a defocus modulation technique. This technique changes special patterns displayed on a spatial light modulator~(SLM) to achieve defocus instead of mechanical translation or focus ring rotation, thus achieve fast capturing and avoid error induced by mechanical movement. We verify the feasibility of the proposed method by two typical focal plane sweeping based light field reconstruction techniques, they are light field moment imaging~(LFMI) and light field reconstruction with back propagation~(LFBP) approach. 

\section{Focal plane sweeping based light field acquisition}

\begin{figure}[htp]
	\captionsetup[subfigure]{justification=centering}
	             \begin{subfigure}[b]{1\linewidth}
	             	\begin{tikzpicture}
				\scope[nodes={inner sep=0,outer sep=0}]
				\node[anchor=south west]
				{
				\centering
			             \begin{overpic}[width=.24\linewidth]{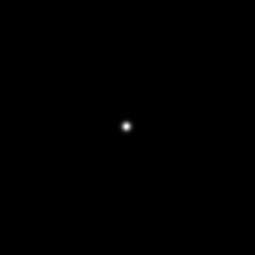}	
			             \put(45, 105){\color{black}{$z_0$}}
				\end{overpic}
				}; 
				\draw[<->,line width=0.2mm, yellow] (0.1,0.4) node (yaxis)[above]{$\textcolor{yellow}{y}$} |- (0.4,0.1) node (xaxis) [right]{$\textcolor{yellow}{x}$};				
			 	\endscope
			\end{tikzpicture}
	             	\begin{overpic}[width=.24\linewidth]{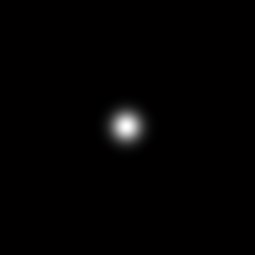}
	             		\put(45, 105){\color{black}{$z_1$}}
	             	\end{overpic}	
	             	\begin{overpic}[width=.24\linewidth]{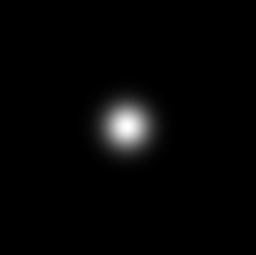}
	             		\put(45, 105){\color{black}{$z_m$}}
	             	\end{overpic}
	             	\begin{overpic}[width=.24\linewidth]{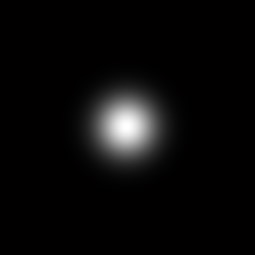}
	             		\put(45, 105){\color{black}{$z_M$}}
	             	\end{overpic}	

	             	\caption{}
	             \end{subfigure}	
	
		\begin{subfigure}[b]{1\linewidth}
			\begin{tikzpicture}
				\scope[nodes={inner sep=0,outer sep=0}]
				\node[anchor=south west]
				{
				\centering
			             \begin{overpic}[width=.24\linewidth]{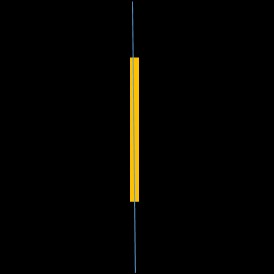}	
				\end{overpic}
				}; 
				\draw[<->,line width=0.2mm, yellow] (0.1,0.4) node (yaxis)[above]{$\textcolor{yellow}{\xi}$} |- (0.4,0.1) node (xaxis) [right]{$\textcolor{yellow}{x}$};
				\draw[thin,-,white] (1,0) -- (1, 2.1) node[anchor=north west]{};
			 	\endscope
			\end{tikzpicture}
			\begin{tikzpicture}
				\scope[nodes={inner sep=0,outer sep=0}]
				\node[anchor=south west]
				{
				\centering
			             \begin{overpic}[width=.24\linewidth]{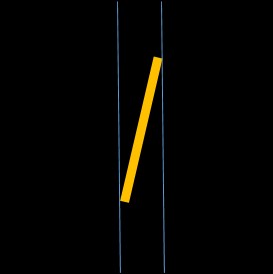}	
				\end{overpic}
				}; 			
				\draw[thin,-,white] (0.9,0) -- (0.9, 2.1) node[anchor=north west]{};
				\draw[thin,-,white] (1.2,0) -- (1.2, 2.1) node[anchor=north west]{};
			 	\endscope
			\end{tikzpicture}
			\begin{tikzpicture}
				\scope[nodes={inner sep=0,outer sep=0}]
				\node[anchor=south west]
				{
				\centering
			             \begin{overpic}[width=.24\linewidth]{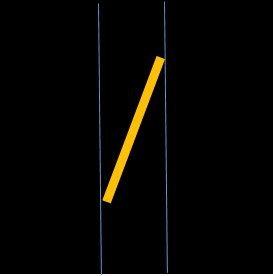}	
				\end{overpic}
				}; 			
				\draw[thin,-,white] (0.78,0) -- (0.78, 2.1) node[anchor=north west]{};
				\draw[thin,-,white] (1.25,0) -- (1.25, 2.1) node[anchor=north west]{};
			 	\endscope
			\end{tikzpicture}
			\begin{tikzpicture}
				\scope[nodes={inner sep=0,outer sep=0}]
				\node[anchor=south west]
				{
				\centering
			             \begin{overpic}[width=.24\linewidth]{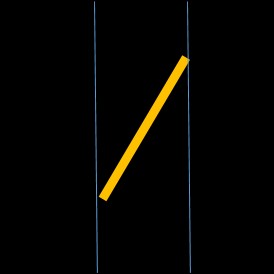}	
				\end{overpic}
				}; 			
				\draw[thin,-,white] (0.73,0) -- (0.73, 2.1) node[anchor=north west]{};
				\draw[thin,-,white] (1.42,0) -- (1.42, 2.1) node[anchor=north west]{};
			 	\endscope
			\end{tikzpicture}
		\caption{}
		\end{subfigure}		

	\caption{(a) Images of a point located at different focal planes of a camera system, and (b) the corresponding EPIs.}
	\label{fig:psf_epi}
\end{figure}
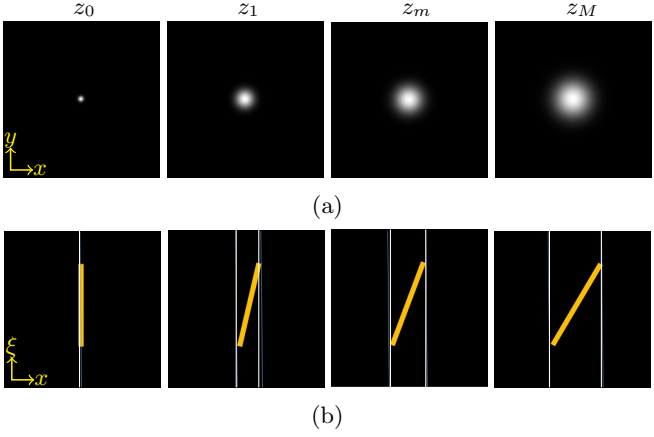

According to the plenoptic function~\cite{levoy2006light}, light field can be parameterized as a five-dimensional function $L(x, y, \xi ,\eta, z )$, where $(x, y, z)$ is the spatial coordinates and $(\xi ,\eta )$ is the angular coordinates.\
In the focal plane sweeping imaging system, suppose $I(x,y,z_m)$ is the $m^{th}$ captured images with the focal plane located at $z_m$, and $M$ is the total number of the captured images. The captured images are the convolution between the clear images and the point spread function~(PSF) of the system~\cite{Park_2010_SPIE,Park_2010_DH}. 
In general, the PSF of a camera can be regarded as Gaussian distribution function due to the circular shape of the optical elements and apertures.  
For a point object, the numerical captured images with focal plane sweeping are shown in Fig.~\ref{fig:psf_epi}~(a). As the definition of PSF,  they equal the 2D slices of the 3D PSF of the camera system. 
Fig.~\ref{fig:psf_epi}~(b) show the corresponding epipolar plane images~(EPI) cross the center horizontal line of the captured images.  Focal plane sweeping in spatial space corresponds shearing of EPI, and the amount of shearing reflects the focal plane sweeping distance. This relationship between the defocused images and the EPIs is the basis of the focal plane sweeping based light field acquisition techniques. In this paper we analyze LFMI and LFBP, which are two typical focal plane sweeping based light field reconstruction techniques. 

\begin{figure}
	\centering
	\includegraphics[width=\columnwidth ]{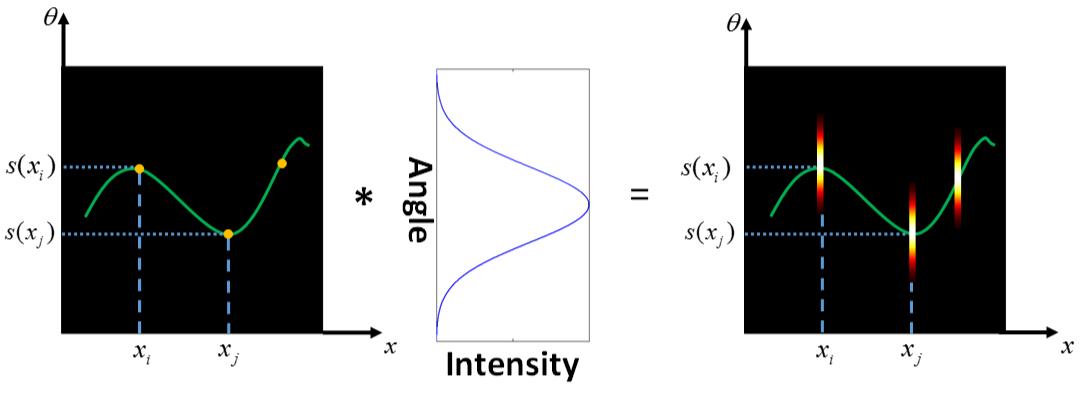}
	\caption{Principle of LFMI represented by EPI. }
	\label{eq:LFMI}
\end{figure}
LFMI constructs an approximate light field at a designed plane ${z_m}$ by an empirical assumption that the angular of the light rays satisfy Gaussian distribution function of standard deviation~\cite{orth2013light}. The Gaussian distribution assumption of the light ray direction comes from the Gaussian PSF of the camera system~\cite{Park_2010_SPIE,Park_2010_DH}. With the light rays' angular moment at each spatial position, the light field can be reconstructed by
\begin{align}
&L(x, y,\xi ,\eta,z_m)  \nonumber \\
=&I(x,y,z_m)\exp \left\{-\frac{[\xi -s(x,y)]^2+[\eta -t(x,y)]^2}{\sigma^2}\right\} \nonumber \\
=&I(x,y,z_m) \delta[\xi-s(x,y), \eta-t(x,y)] * G(\xi, \eta, \sigma),
\label{eq:recLF}
\end{align}
where $[s(x,y), t(x,y)]$ is the first order angular moment of the light ray at position of $(x, y, z_m)$, $G(\xi, \eta, \sigma)$ is the Gaussian distribution function, $\sigma$ equals the numerical aperture~(NA) of the camera, and $*$ is convolution operator. This can be seen more intuitively from Fig.~\ref{eq:LFMI}. The estimated angular moment is a sparse sampling of the EPI, as the left image in Fig.~\ref{eq:LFMI} shows, $s(x)$ is the angular moment at position $(x)$, which is the average light ray direction. The final calculated EPI~(Right image in Fig.~\ref{eq:LFMI}) is the convolution between the angular moment and the Gaussian PSF~(Center image in Fig.~\ref{eq:LFMI}). It can be seen that the final EPI is mainly determined by the angular moment, therefore its accuracy affects the reconstructed light field most importantly. 
In LFMI, it has been proved the light ray transport along the the optical axis satisfies a partial differential equation~(PDE), and the angular moment is acquired by solving this PDE. It is obvious that the quantity of light ray transport depends on both the depth interval of the images and the bandwidth of the object very much. Therefore, the depth interval between two adjacent defocused images should be choosn carefully according to the object's characteristics~\cite{orth2013light}. In general, a conventional camera system's PSF is determined, and the light transport can only be controlled by the depth interval of the captured images, this makes it difficult to apply LFMI to an specific object. 
Usually, at least two defocused images works for determining the light transport, but with a large stack of images, we can estimate high order angular moment, thus calculate more accurate light field~\cite{liu2015light}. 
With the above analysis, we can improve the LFMI in two aspects, one is capturing more focal plane sweeping images, and the other one is designing a focal plane sweeping imaging system with a controllable PSF. 

\begin{figure}[htp]
	\captionsetup[subfigure]{justification=centering}
	\begin{subfigure}[b]{1\linewidth}
		\begin{tikzpicture}
			\scope[nodes={inner sep=0,outer sep=0}]
			\node[anchor=south west]
			{
			\centering
		             \begin{overpic}[width=.24\linewidth]{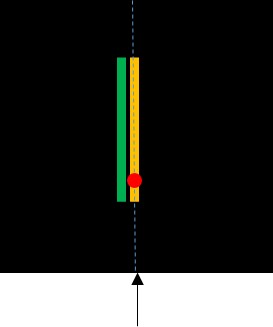}	
		              \put(35, -10){\color{black}{$x_0$}}
		              \put(35, 110){\color{black}{$z_0$}}
			\end{overpic}
			}; 
			\draw[<->,line width=0.2mm, yellow] (0.1,0.8) node (yaxis)[above]{$\textcolor{yellow}{\xi}$} |- (0.5,0.5) node (xaxis) [right]{$\textcolor{yellow}{x}$};
			\draw[thin,-,white] (1,0.4) -- (1, 2.5) node[anchor=north west]{};
		 	\endscope
		\end{tikzpicture}
		\begin{tikzpicture}
			\scope[nodes={inner sep=0,outer sep=0}]
			\node[anchor=south west]
			{
				\centering
			             \begin{overpic}[width=.24\linewidth]{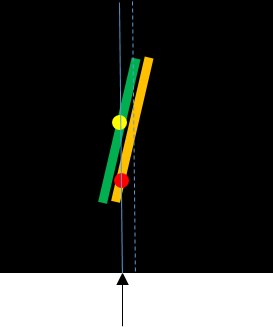}	
			             \put(18, -10){\color{black}{$ x_1=\xi_0 z_1$}}
			             \put(35, 110){\color{black}{$z_1$}}
				\end{overpic}
			}; 			
			\draw[thin,-,dashed,white] (0.9,0.4) -- (0.9, 2.5) node[anchor=north west]{};
			\draw[thin,-,white] (1,0.4) -- (1, 2.5) node[anchor=north west]{};
			 \endscope
		\end{tikzpicture}
		\begin{tikzpicture}
			\scope[nodes={inner sep=0,outer sep=0}]
			\node[anchor=south west]
			{
				\centering
			             \begin{overpic}[width=.24\linewidth]{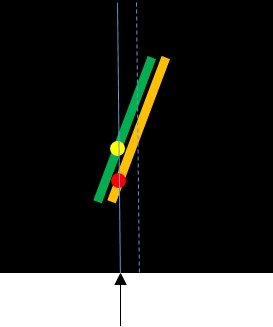}	
			             \put(15, -10){\color{black}{$ x_m=\xi_0  z_m$}}
			             \put(35, 110){\color{black}{$z_m$}}
				\end{overpic}
			}; 			
			\draw[thin,-,dashed,white] (0.9,0.4) -- (0.9, 2.5) node[anchor=north west]{};
			\draw[thin,-,white] (1.05,0.4) -- (1.05, 2.5) node[anchor=north west]{};
			 \endscope
		\end{tikzpicture}
		\begin{tikzpicture}
			\scope[nodes={inner sep=0,outer sep=0}]
			\node[anchor=south west]
			{
				\centering
			             \begin{overpic}[width=.24\linewidth]{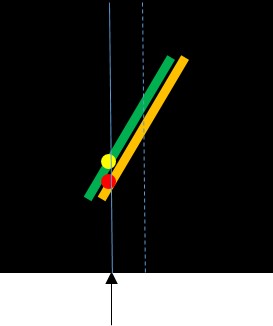}	
			             \put(15, -10){\color{black}{$ x_M=\xi_0  z_M$}}
			             \put(35, 110){\color{black}{$z_M$}}
				\end{overpic}
			}; 			
			\draw[thin,-,dashed,white] (0.83,0.4) -- (0.83, 2.5) node[anchor=north west]{};
			\draw[thin,-,white] (1.05,0.4) -- (1.05, 2.5) node[anchor=north west]{};
			 \endscope
		\end{tikzpicture}
	\end{subfigure}
	\vspace{1pt}
	\caption{Principle of LFBP represented by EPI.}
	\label{fig:LFBP}
\end{figure}
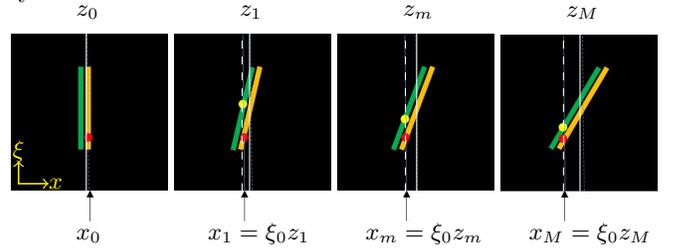
In LFBP,  the light field with the principal plane located at $z = 0$ is calculated by~\cite{park2014light}:
\begin{equation}\label{eq:LFBP}
	L(x,y,\xi,\eta, z_0)=\sum_{m=1}^{M}I\left( \frac{x+z_m\xi}{\alpha},\frac{y+z_m\eta}{\alpha}, z_m \right),
\end{equation}
where $\alpha$ is the magnification between the camera sensor plane and the focal plane. The description of the principle is represented by Fig.~\ref{fig:LFBP} more intuitively. As previously description in Fig.~\ref{fig:psf_epi}, focal plane sweeping in spatial space induces shearing in the light field space. For a given spatial position and a specific light ray direction, the spatial positions that the light ray goes through at each defocused image are determined, as the horizontal shift of red points in each image of Fig.~\ref{fig:LFBP}. The first image represents the EPI corresponding to a focal plane at $z_0$. The red point in the first image represents the light field of $L(x_0, \xi_0)$, the corresponding position at the other focal planes is $x_m=\xi_0 z_m$, as the dashed white lines show. Therefore, the radiance of the specific light ray can be obtained by averaging the corresponding radiance from all of the defocused images. 
For a real scene, the radiance of each point on the captured images is the the accumulation of all light rays reach at it with different directions, this induces defocus noise in LFBP. As the green lines and yellow points show in Fig.~\ref{fig:LFBP}. The green lines represent a point at the same depth as the yellow lines represented, but with a different lateral position. The intensity at $x_m$ is the integral along the white dashed lines. It can been seen that the yellow points from the green lines also contribute to the intensities. When we reconstruct the light field at a specific point, much noise from all the other points is induced. Fortunately, the red point change position in a linear transformation and the noise from all the other points is different at different defocus length. By summing all image with linear transform, the actual light field have the largest weight. 
It's obvious that  with a large camera NA, the defocus noise from all the other points can be reduced because the summing weight of the noise will be reduced. Besides, It has been proved that the depth resolution of the reconstructed light field depends on the depth interval of the captured images, i.e., more defocused images achieve better depth resolution~\cite{Chen_2017_AO}.

From the previous contents, we can see that PSF of the camera system that used for capturing the focal sweeping images is critical in the light field reconstruction. In LFMI, it affects the accuracy of the calculated light angular moment as well as the light field. In LFBP, it is a critical factor that affects the defocus noise in the reconstructed light field. Further more, in both of the two techniques, more focal plane sweeping images achieve better light field reconstruction. In LFMI, higher order angular moment can be obtained from more images, and in LFBP, more images achieves higher axial resolution. However, more focal plane captured images is time consuming and induce alignment and magnification problems~\cite{Park_2010_SPIE,Park_2010_DH}. Therefore, controlling the PSF of the focal plane sweeping imaging system is of great importance. Actually, PSF of an imaging system can be manipulated for many applications, this is called PSF engineer in many other research fields~\cite{Chen_2016_PR}. In this paper we insert a PSF modulation component into a conventional microscopic imaging system. On one hand, this achieves accelerated speed and more accurate focal plane sweeping capture. One the other hand, more freely PSF control can be performed for specific requirements. In the following section, we describe how we manipulate the PSF of the imaging system to achieve a focal plane sweeping image capture without translation movement of the camera or the object. 

\section{Focal plane sweeping with defocus modulation}
\begin{figure} [!t]
	\centering
	\captionsetup[subfigure]{justification=centering}
	\includegraphics[width=0.75\columnwidth]{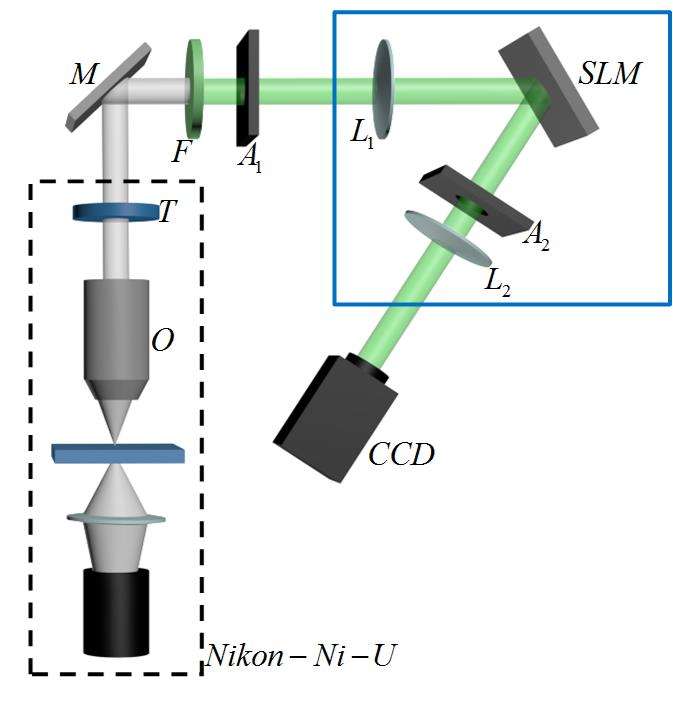}
	\caption{The schematic of the experimental setup. $M$ is a mirror, $F$ is a light filter, $A_1, A_2$ are apertures, $L_1, L_2$ are thin lens.}
	\label{fig:setup}
\end{figure}

The setup scheme of our proposed system is shown in Fig.~\ref{fig:setup}. The components within the dashed rectangle is a commercial microscope~(Nikon Ni-U). A mirror~(M) is used to export the light from the microscope. $F$ is a light filter with a bandwidth of $\SI{3}{\nano\meter}$ at the wavelength of $\SI{532}{\nano\meter}$. An aperture $A_1$ is located at the imaging plane of the microscope, which is used for adjusting image size. The other aperture $A_2$ is used for selecting the first diffraction order of the SLM. The components within the solid rectangle is used for PSF modulation. Lenses $L_1$ and $L_2$ form the $4f$ system.  An SLM~(Holoeye, LETO) is located at the Fourier spectrum plane of the $4f$ system, which performs the PSF modulation. The SLM is a phase-only modulator, which transforms phase shift in a range of  $[0, 2\pi]$ to 8-bit gray levels. The CCD~(PointGrey, GS3-U3-23S6M-C) plane is conjugated with the image plane of the microscope. In the following paragraphs we explain how we control the patterns in the SLM to achieve PSF modulation and analyze the performance of it.

\subsection{Principle of the PSF modulation}
In our system, the SLM acts as a Fresnel lens with a desired focal length. The modification of the focal length on the SLM produces a focal plane sweeping, thus making the captured images equals captured at different depths. Suppose a desired corresponding axial focal plane shift of $z_i$ in the imaging plane is required,  the modulation focal length of the SLM should be~\cite{maurer2010depth,djidel2006high}:
\begin{equation}\label{eq:f_slm}
	f_{SLM}=-\frac{f_r^2}{z_i},
\end{equation}
where $f_r$ is the focal length of lens $L_1 $. The axial shift at the sample stage is $z_o = z_i/\beta^2$, and $\beta$ is the magnification of the objective. The required phase pattern displayed on the SLM thus can be written as~\cite{maurer2010depth}:
\begin{align}\label{eq:phi}
	\varphi(x,y) \nonumber = &\frac{\pi}{\lambda f_{SLM}}(x^2+y^2) \nonumber \\
	=&-\frac{\pi  z_i}{\lambda f_r^2}(x^2+y^2),
\end{align}
where $\lambda$ is the light wavelength and $(x, y)$ are the spatial coordinates. 

\subsection{Defocus performance of the proposed system}
It should be noted that the SLM is pixellated and the phase represented by the SLM is discrete. Therefore the corresponding depth range and depth interval that can be modulated by our system are limited. Here analyze the two limitations and give the two values according to the specifications of our system.

Because of the pixellated SLM, the phase that can be represented by the SLM is limited by~\cite{maurer2010depth}
\begin{equation}\label{eq:dphi}
	|\Delta \varphi |< \pi,
\end{equation}
where $p$ is the pixel pitch of the SLM. This results in a limited corresponding depth range that can be represented by the proposed system. Substituting Eq.~(\ref{eq:dphi}) into Eq.~(\ref{eq:phi}), we obtain the maximum depth shift that can be represented according to the system specifications
\begin{equation}\label{eq:z_max}
	|z_{max}|=\frac{\lambda f_r^2}{2p r_{l}},
\end{equation}
where $r_{l}$ is the radius of the light enter into the SLM. Generally, we let $r_{l}\le0.5\min(x_{max},y_{max})$, this make sure that the light is within the effective area of the SLM. $(x_{max},y_{max})$ are the length and width of the SLM. The center of the SLM is coincide with the optical axis, making the lateral position of the images on the CCD remain changeless. 

Since the gray level that represented by the SLM is 8-bit, which corresponds to a discrete phase value, the minimal phase change on the SLM is $\varphi_{min} = {2 \pi}/{256}$. Suppose the corresponding minimal depth change is $\Delta z_{min}$, from Eq.~(\ref{eq:phi}) we get
\begin{equation}\label{eq:dz}
	|\Delta z_{min}| =\frac{\lambda f_r^2}{128r^2}.
\end{equation}

In our experimental setup, we used a $20\times$ objective. The other parameters are $p=\SI{6.6}{\micro\meter}$, $\lambda = \SI{532}{\nano\meter}$, $r_l = 0.5 x_{max}\simeq \SI{3.3}{\milli\meter}$, $f_r=\SI{200}{\milli\meter}$. From Eq.~(\ref{eq:z_max}) and Eq.~(\ref{eq:dz}), the maximum defocused depth is $\SI{488.5}{\milli\meter}$ and the minimum defocus depth shift is $\SI{0.0152}{\milli\meter}$.  In the experiment, according to the light field reconstruction technique, we can control the PSF by choosing proper patterns to be displayed on the SLM, but we should make sure the phase patterns on the SLM satisfy the two limits.

In addition to the above limits, it is worth to mention that the bandwidth of the light filter has a great influence on the image quality due to the single wavelength selection of the SLM.
The patterns on the SLM require additional grating phase to separate the modulated and unmodulated light, but the grating phase would lead to distinct dispersion. The bandwidth of the light filter should be narrow enough, which also induces light attenuation. 
Besides, the SLM is not located at the exact Fourier plane of lens $L_1$, while it is located on the imaging plane of the collector lens. We can see distinct images of the dot on the collector lens as well as the edge of the condenser aperture diaphragm. Only in this plane can the magnification of the recored images remain unchanged when we change the focal length of the patterns displayed on SLM.
Furthermore, In order to avoid influence from the previous patterns, we should control the SLM and CCD sequentially to capture the images at each focal plane. 

\section{Experimental results and discussion}
We verify the feasibility and possibility of  light field reconstruction with the proposed imaging system in the following sections. 

\subsection{Verification of PSF modulation}
\begin{figure} [htb]
             \centering
             \captionsetup[subfigure]{justification=centering}
             \begin{subfigure}[b]{1\linewidth}
             \begin{tikzpicture}
			\scope[nodes={inner sep=0,outer sep=0}]
			\node[anchor=south west]
			{
			            \begin{overpic}[width=.18\linewidth]{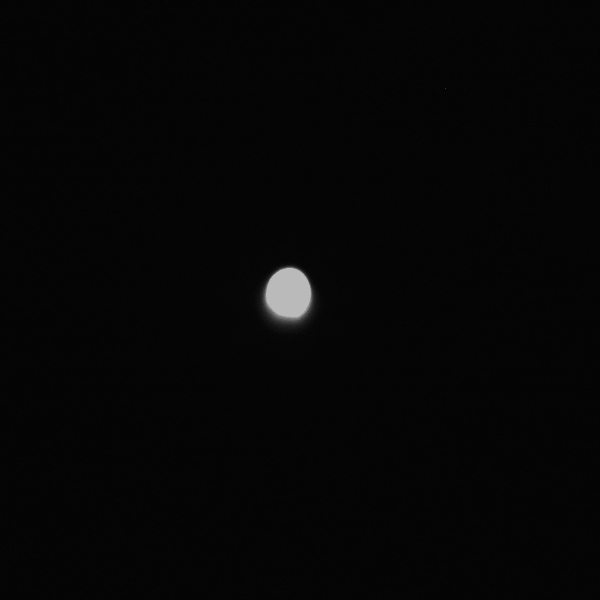}
				\put(2, 105){\color{black}{$z= \SI{0}{\milli\meter}$}}
				\end{overpic}
				 \begin{overpic}[width=.18\linewidth]{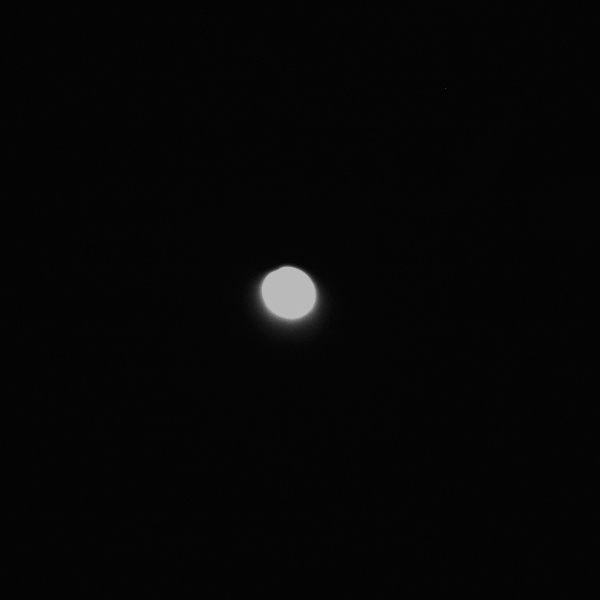}
				\put(5, 105){\color{black}{$z= \SI{10}{\milli\meter}$}}
				\end{overpic}	
				 \begin{overpic}[width=.18\linewidth]{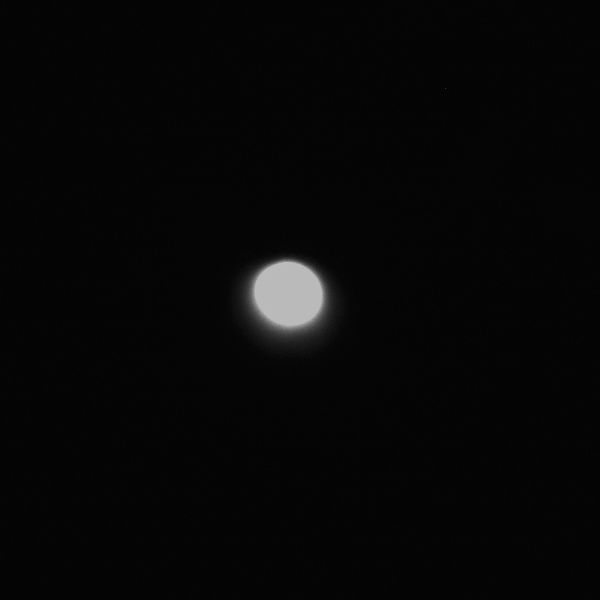}
				\put(5, 105){\color{black}{$z= \SI{20}{\milli\meter}$}}
				\end{overpic}
				 \begin{overpic}[width=.18\linewidth]{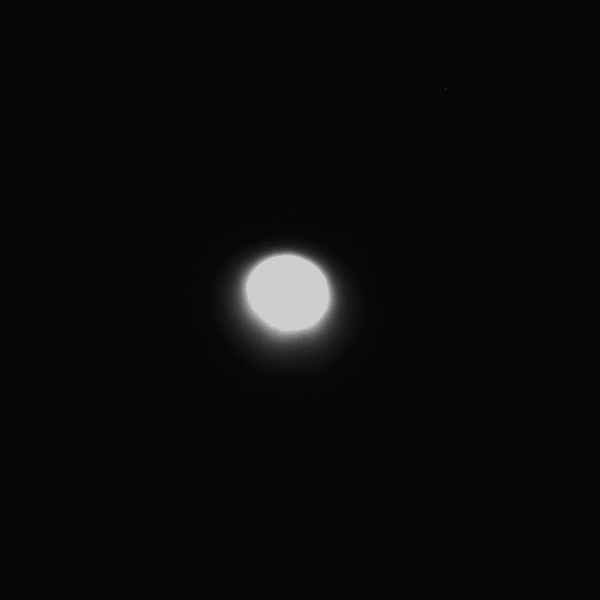}
				\put(5, 105){\color{black}{$z= \SI{30}{\milli\meter}$}}
				\end{overpic}	
			            \begin{overpic}[width=.18\linewidth]{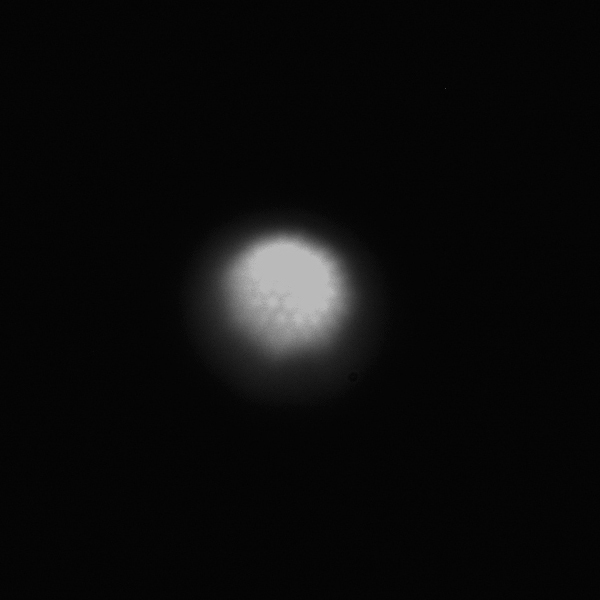}
				\put(5, 105){\color{black}{$z= \SI{40}{\milli\meter}$}}
				\end{overpic}	
			}; 			
			\endscope
	\end{tikzpicture}
             \caption{}
	\end{subfigure}	

             \begin{subfigure}[b]{1\linewidth}
	\begin{tikzpicture}
			\scope[nodes={inner sep=0,outer sep=0}]
			\node[anchor=south west]
			{
			             \begin{overpic}[width=.18\linewidth]{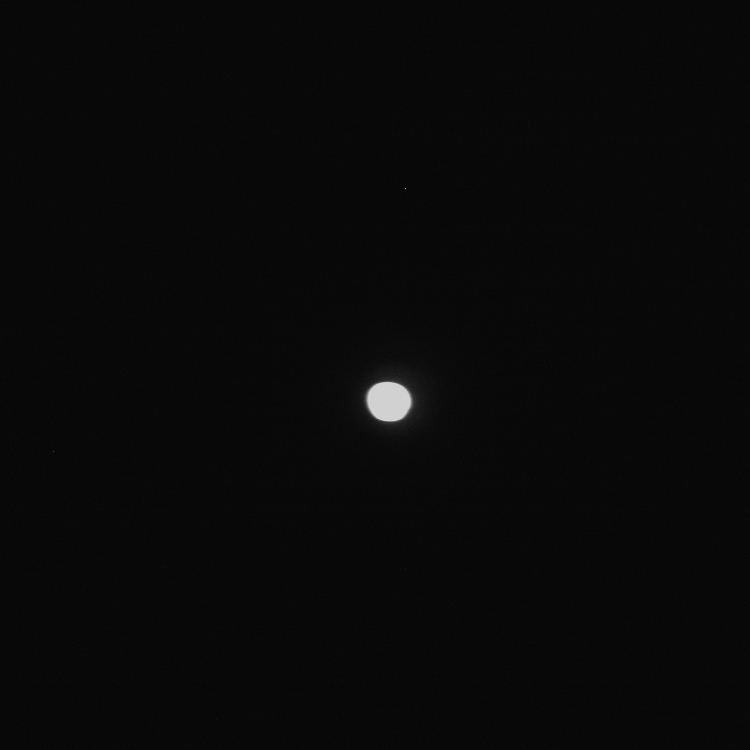}	
				\end{overpic}
			             \begin{overpic}[width=.18\linewidth]{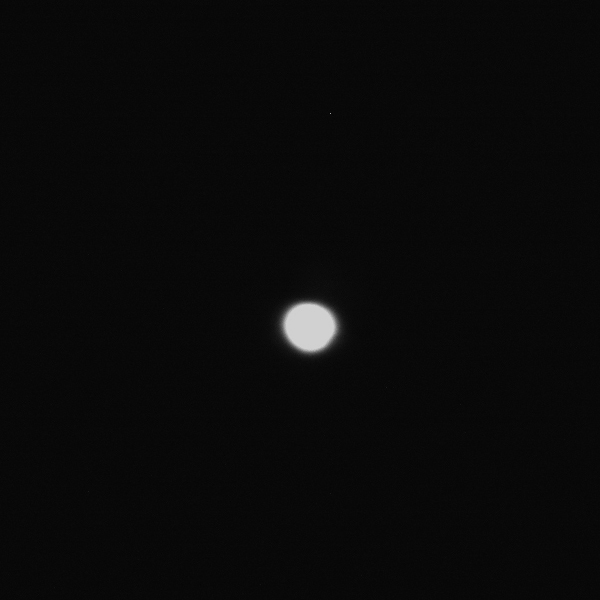}
				\end{overpic}	
			             \begin{overpic}[width=.18\linewidth]{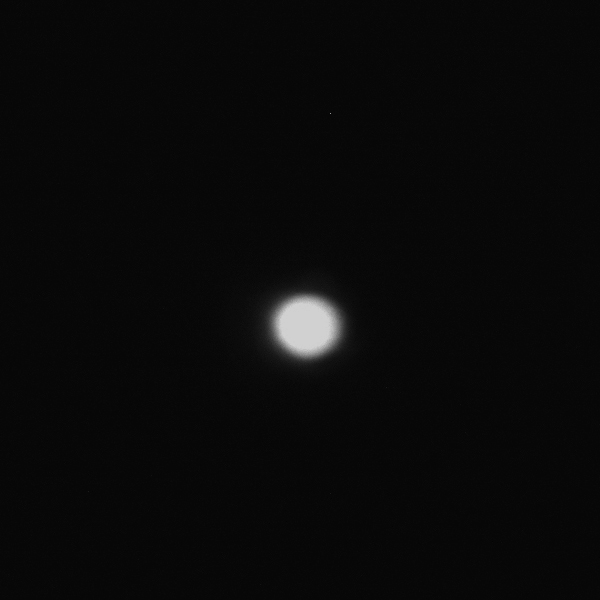}
				\end{overpic}	
				 \begin{overpic}[width=.18\linewidth]{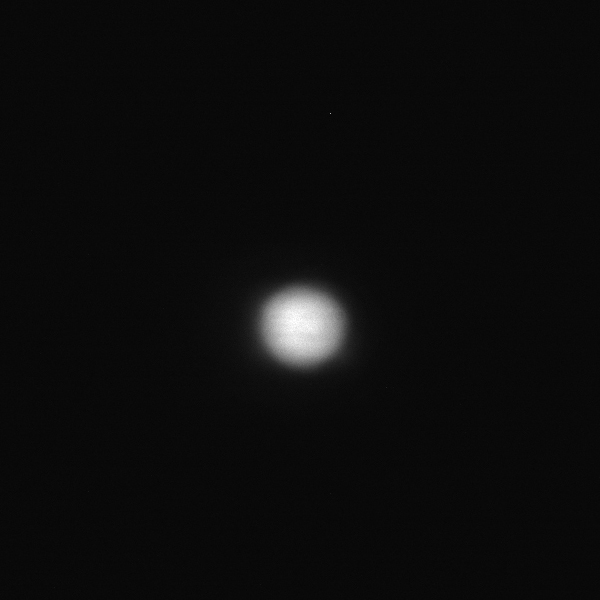}
				\end{overpic}
				 \begin{overpic}[width=.18\linewidth]{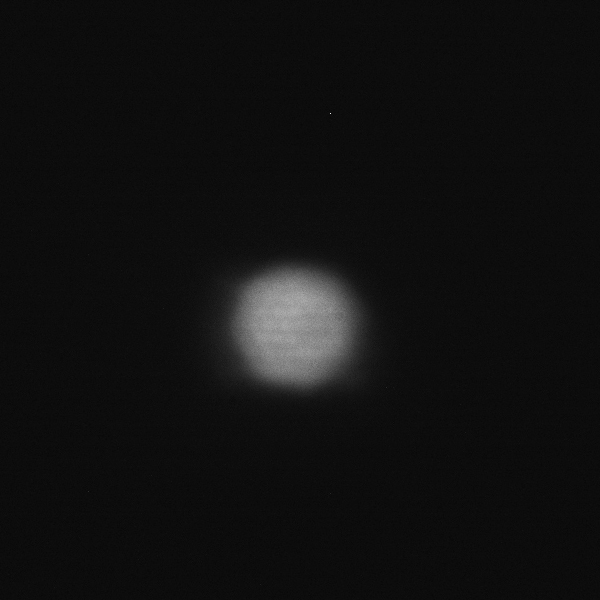}
				\end{overpic}	
			}; 
			\draw[thin, -, white] (0.1,0.2) -- (0.18, 0.2) node[right]{ \tiny{$ \SI{200}{\micro\meter}$}};
			 \endscope
	\end{tikzpicture}

             \caption{}
	\end{subfigure}	
	\caption{The captured images of the PSF at several focal depths with (a) the conventional translation system and  (b) the proposed PSF modulation system, respectively.}
	\label{fig:sys_psf}
\end{figure}

With the proposed system described in the previous section, we have captured the PSF images at several depths, as shown in Fig.~\ref{fig:sys_psf}(b). A pinhole with a diameter of $\SI{10}{\micro\meter}$ was used as a point object. The images captured with the conventional translation system were captured as the ground truth, as shown in Fig.~\ref{fig:sys_psf}(a).  We can observe that the PSF of the proposed system coincides with the the one of the conventional system. This can also be verified by the size of the PSFs. The objective is $20\times$,  the pixel pitch of the captured images is $\SI{5.86}{\micro\meter}$. With the two parameters, all the size of the PSFs can be calculated and verified. The expected diameters of the PSF at the five axial positions should be [$\SI{200}{\micro\meter}$, $\SI{300}{\micro\meter}$, $\SI{400}{\micro\meter}$, $\SI{500}{\micro\meter}$, $\SI{600}{\micro\meter}$]. The measured diameters of the PSF captured with the conventional and the proposed systems are [$\SI{302.1}{\micro\meter}$, $\SI{343.9}{\micro\meter}$, $\SI{444.5}{\micro\meter}$, $\SI{551.4}{\micro\meter}$, $\SI{727.6}{\micro\meter}]$ and $[\SI{257.7}{\micro\meter}$, $\SI{317.1}{\micro\meter}$, $\SI{391.4}{\micro\meter}$, $\SI{456.0}{\micro\meter}$, $\SI{629.8}{\micro\meter}$]. The results show that the PSF of the proposed system is closer to the Gaussian distribution function than the conventional one.  Besides, the shape of the PSF images captured by the proposed system are more likely be circle than the conventional one.

\begin{figure} [htb]
	\captionsetup[subfigure]{justification=centering}
	\vspace{6pt}
	\begin{subfigure}[b]{1\linewidth}
	\begin{tikzpicture}
	\scope[nodes={inner sep=0,outer sep=0}]
	\node[anchor=south west]
	{
	\centering
             \begin{overpic}[width=.45\linewidth]{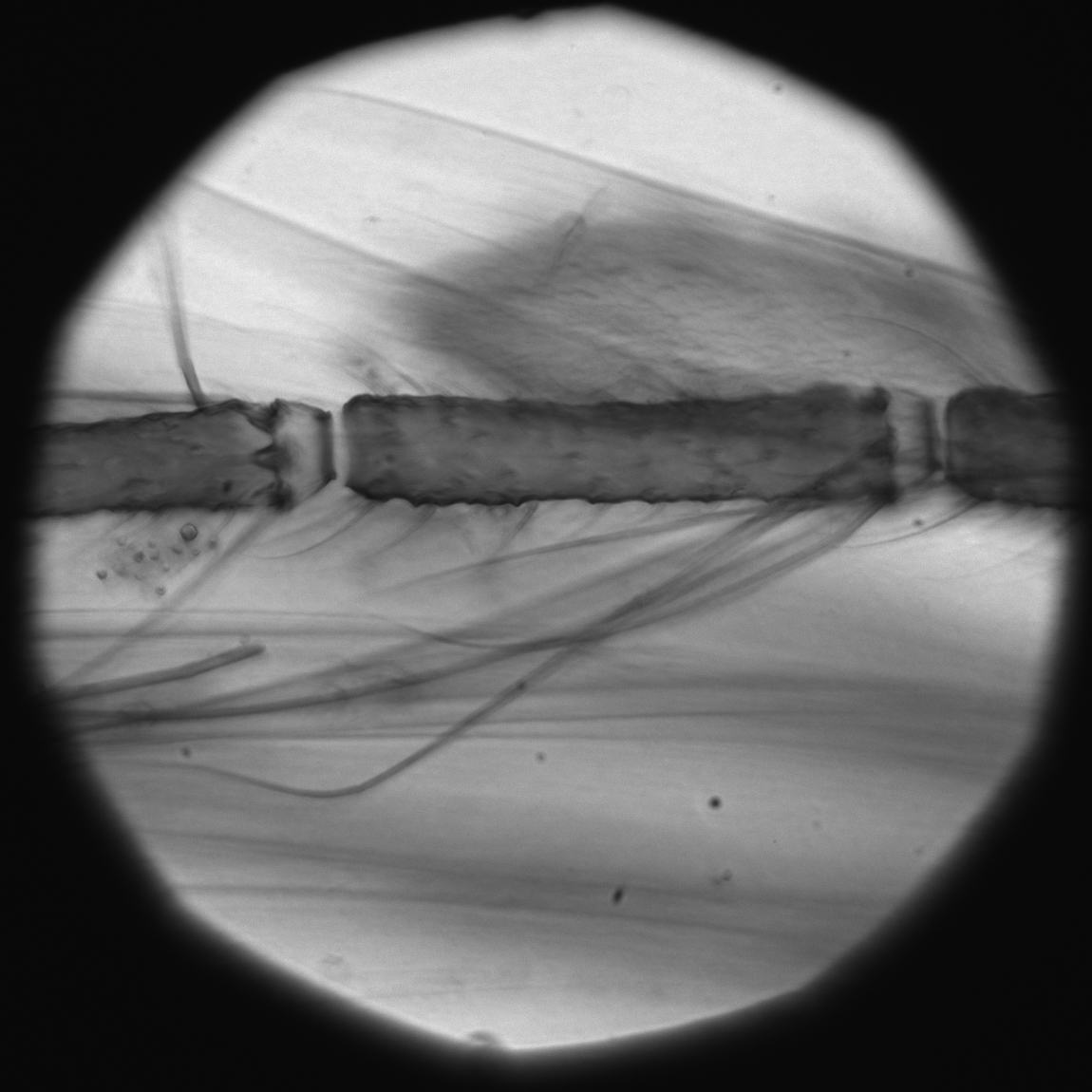}
	\put(35, 105){\color{black}{$z= \SI{11}{\milli\meter}$}}
	\end{overpic}
	 \begin{overpic}[width=.45\linewidth]{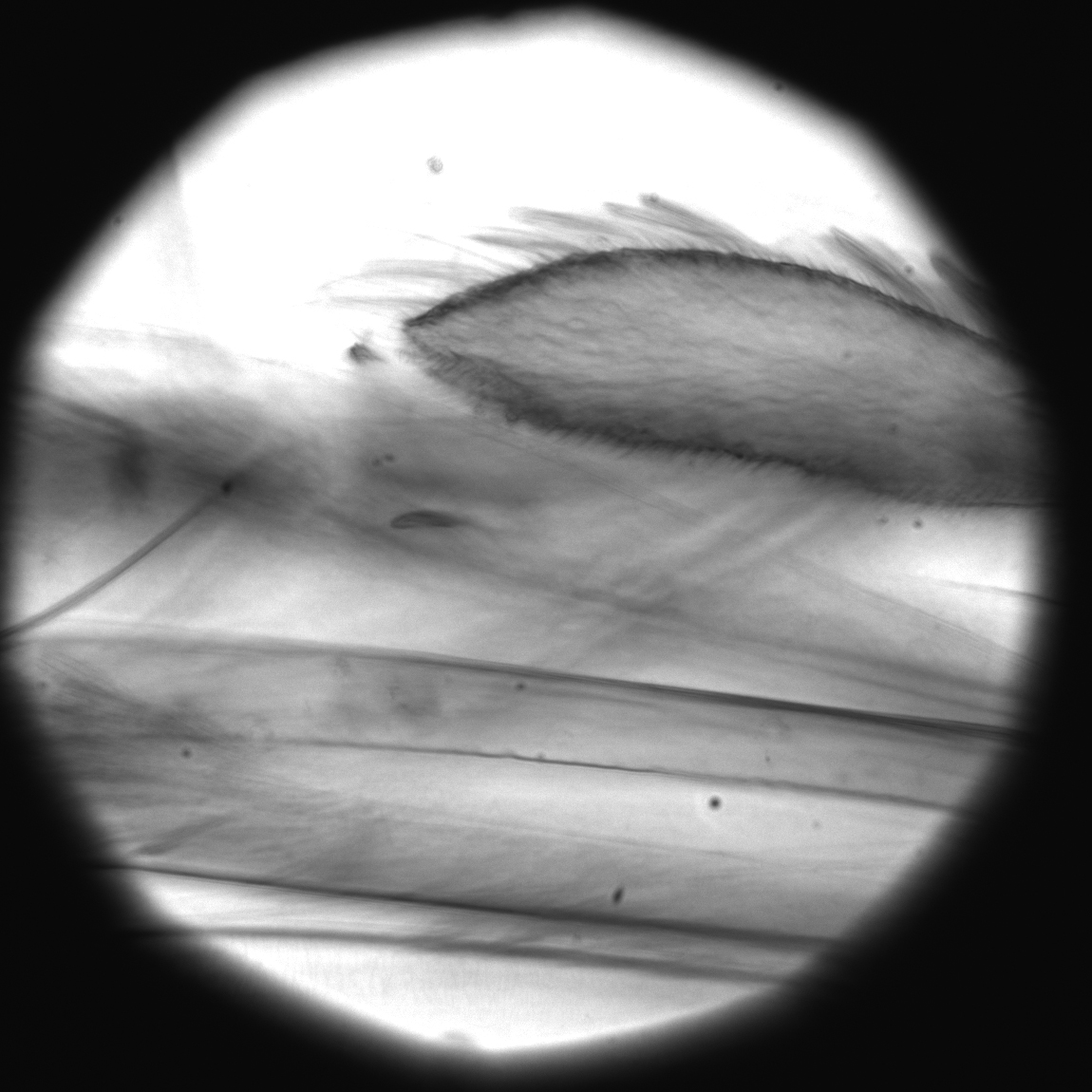}
	\put(35, 105){\color{black}{$z= \SI{60}{\milli\meter}$}}
	\end{overpic}
	}; 
	\draw[-,line width=0.3mm, yellow] (3.75, 0) -- (3.75, 2.5);	
	\draw[-,line width=0.3mm, yellow] (7.6, 0) -- (7.6, 2.5);	
 	\endscope
	\end{tikzpicture}
	\caption{}
	\end{subfigure}		
             
             \vspace{10pt}
             \begin{subfigure}[b]{1\linewidth}
	\begin{tikzpicture}
	\scope[nodes={inner sep=0,outer sep=0}]
	\node[anchor=south west]
	{
	\centering
             \begin{overpic}[width=.45\linewidth]{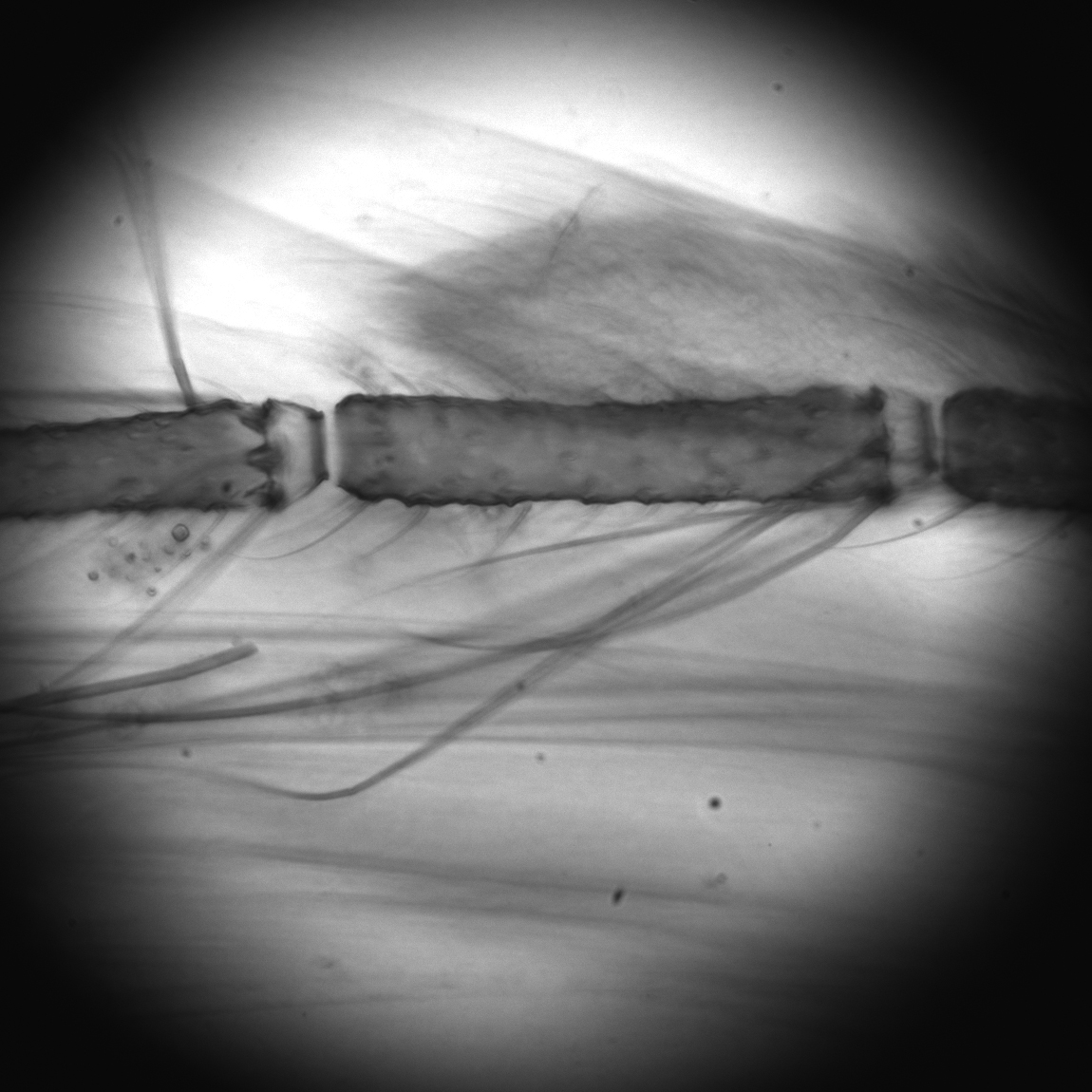}
	\put(35, 105){\color{black}{$z= \SI{11}{\milli\meter}$}}
	\end{overpic}
	 \begin{overpic}[width=.45\linewidth]{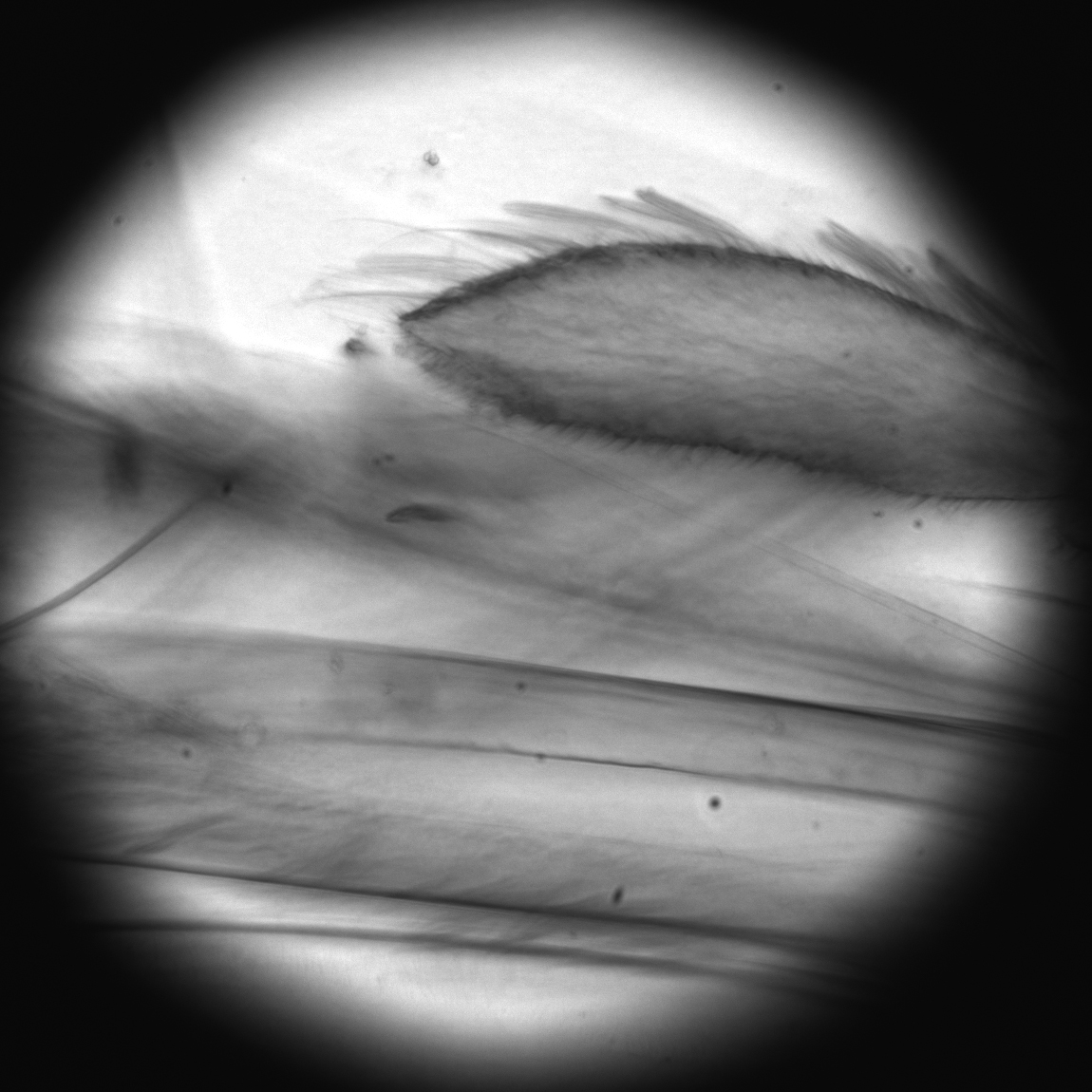}
	\put(35, 105){\color{black}{$z= \SI{60}{\milli\meter}$}}
	\end{overpic}
	}; 
 	\endscope
	\end{tikzpicture}
	\caption{}
	\end{subfigure}		

	\caption{The captured images at two focal planes with (a) translation movement and (b) PSF modulation, respectively.}
	\label{fig:capture_img}
\end{figure}
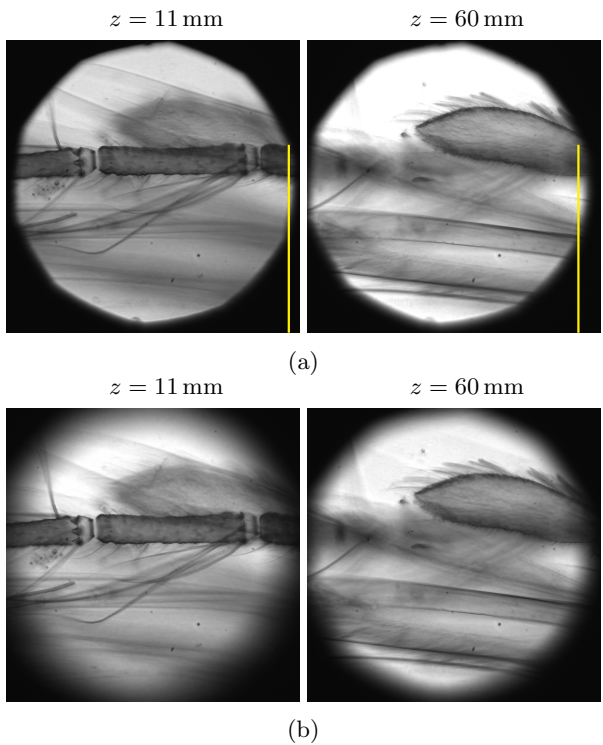

Fig.~\ref{fig:capture_img} shows some captured images with the conventional and proposed systems. Fig.~\ref{fig:capture_img}(a) are the images captured by the conventional translation movement system at $z=\SI{11}{\milli\meter}$ and $z=\SI{600}{\milli\meter}$, respectively. Fig.~\ref{fig:capture_img}(b) show the images captured at the same axial positions with the proposed system. We can observe that the focal plane sweeping capture can really be achieved with our proposed system. However, even through we have calibrated the system very carefully, lateral shift can be clearly observed from the aperture shift in Fig.~\ref{fig:capture_img}(a), as the yellow lines show. 

We have also compared the capture time between the conventional and the proposed systems. 
$61$ image were captured with the two systems, it costs about $30$ minutes and $25$ seconds respectively. It should be mentioned that in the capture process, all the translation movement and the pattern modification on the SLM were manually operated. Reduced time requirement is expected by computational controlling of the systems, but the problems induced by movement in the conventional system still exist, and the translation is still more time consuming than PSF modulation. 

\subsection{Light field reconstruction from focal plane sweeping captured images with PSF modulation}
\begin{figure}[htp]
	\captionsetup[subfigure]{justification=centering}
	\begin{subfigure}[b]{0.45\linewidth}
	\centering
	\includegraphics[width=1\columnwidth]{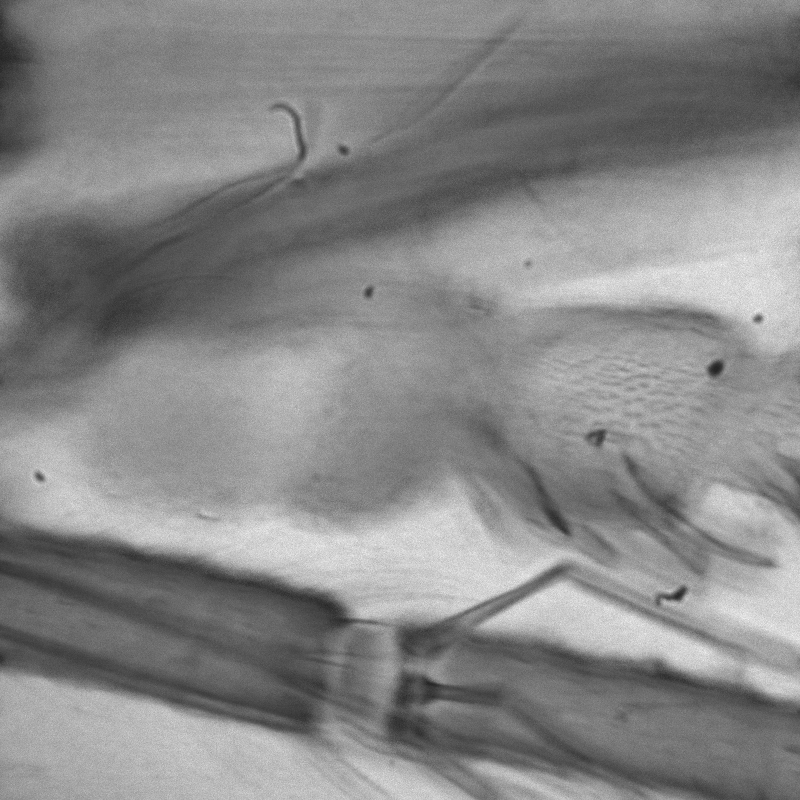}
	\caption{}
	\end{subfigure}		
	\begin{subfigure}[b]{0.45\linewidth}
	\centering
	\includegraphics[width=1\columnwidth]{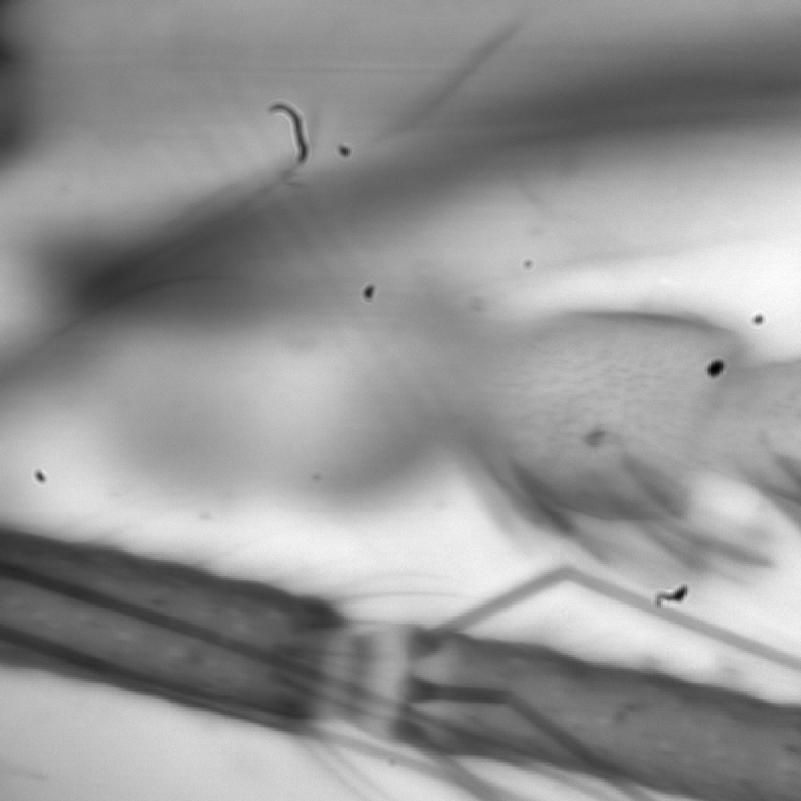}
	\caption{}
	\end{subfigure}

	\begin{subfigure}[b]{0.45\linewidth}
	\centering
	\includegraphics[width=1\columnwidth]{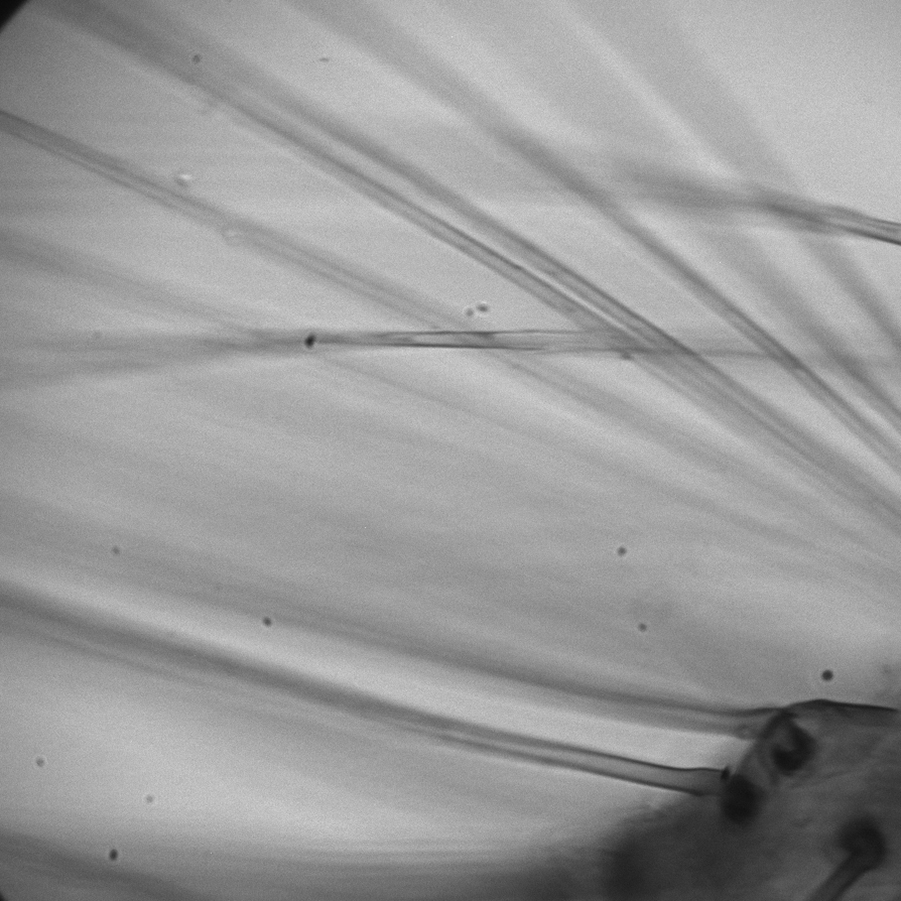}
	\caption{}
	\end{subfigure}		
	\begin{subfigure}[b]{0.45\linewidth}
	\centering
	\includegraphics[width=1\columnwidth]{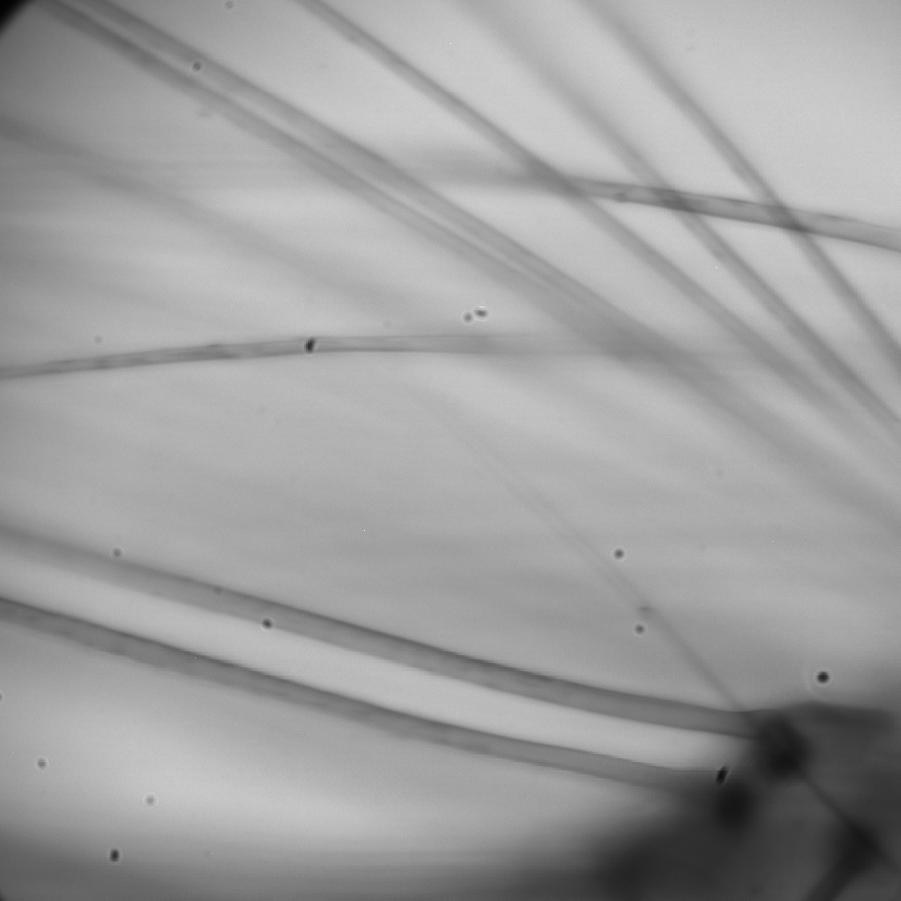}
	\caption{}
	\end{subfigure}

	\caption{Reconstructed parallax view images with (a)(c) LFMI~(see \href{Visualization1.avi}{Visualization 1} and  \href{Visualization3.avi}{Visualization 3}) and (b)(d)~LFBP~(see \href{Visualization2.avi}{Visualization 2} and \href{Visualization4.avi}{Visualization 4})  for mosquito's mouth and mosquito larva.}
	\label{fig:LF}
\end{figure}
We have also verified the light field reconstruction with the two systems. Two objects were used to perform the light filed reconstruction from the captured images. 
We have captured a stack of intensity images of the sample with a corresponding axial spacing of $\Delta z = \SI{1}{\micro\meter}$. 60 defocused images were captured for each object. And 11 images were used for the light field reconstruction. Fig.~\ref{fig:LF}(a)(c) and (b)(d) are the reconstructed parallax view images with the LFMI and LFBP respectively. While Fig.~\ref{fig:LF}(a)(b) are the images of  mosquito's mouth, and (b)(d) are the images of mosquito larva. More parallax view images can be observed from the videos. Both objects were reconstructed with clear parallax with the two light field reconstruction techniques.

\begin{figure}[!t]
	\captionsetup[subfigure]{justification=centering}
	\begin{subfigure}[b]{0.45\linewidth}
	\centering
	\includegraphics[width=1\columnwidth]{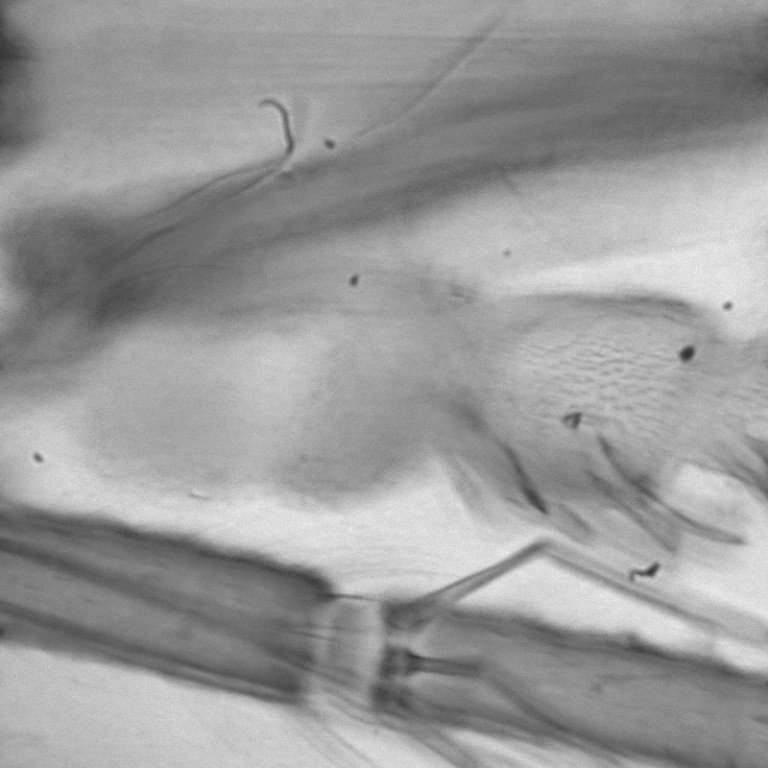}
	\caption{}
	\end{subfigure}		
	\begin{subfigure}[b]{0.45\linewidth}
	\centering
	\includegraphics[width=1\columnwidth]{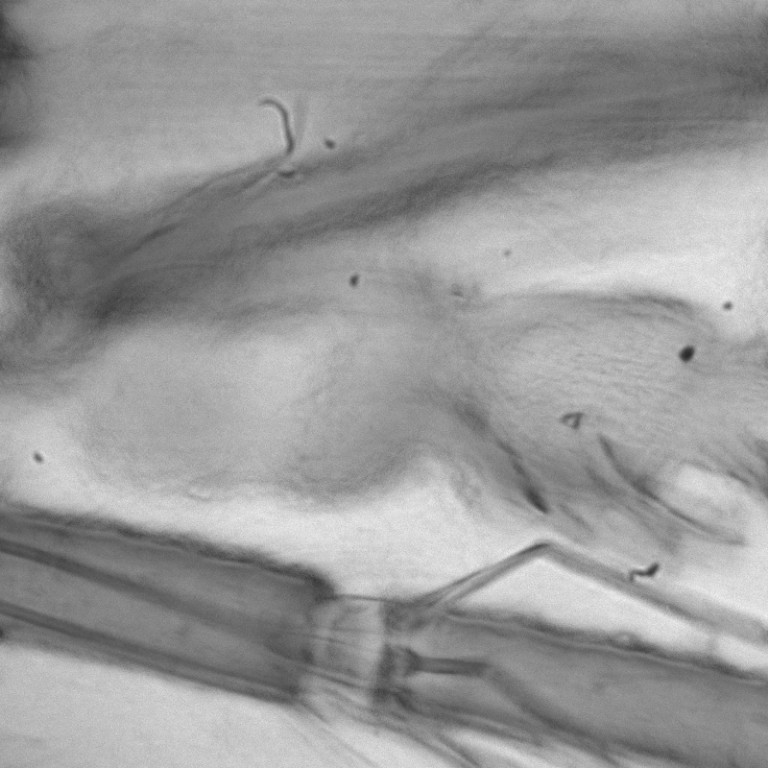}
	\caption{}
	\end{subfigure}

	\begin{subfigure}[b]{0.45\linewidth}
	\centering
	\includegraphics[width=1\columnwidth]{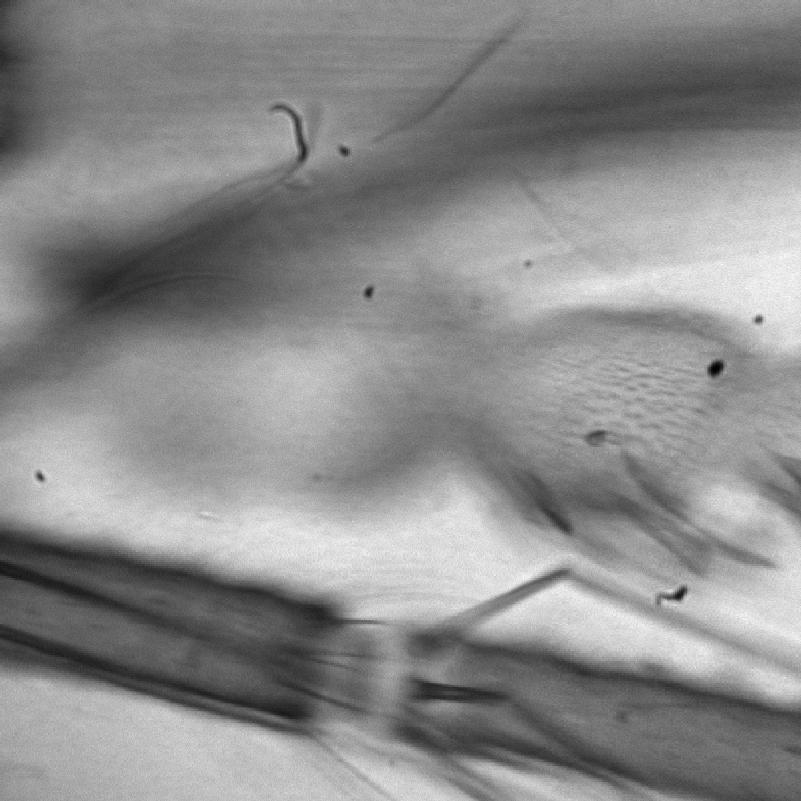}
	\caption{}
	\end{subfigure}		
	\begin{subfigure}[b]{0.45\linewidth}
	\centering
	\includegraphics[width=1\columnwidth]{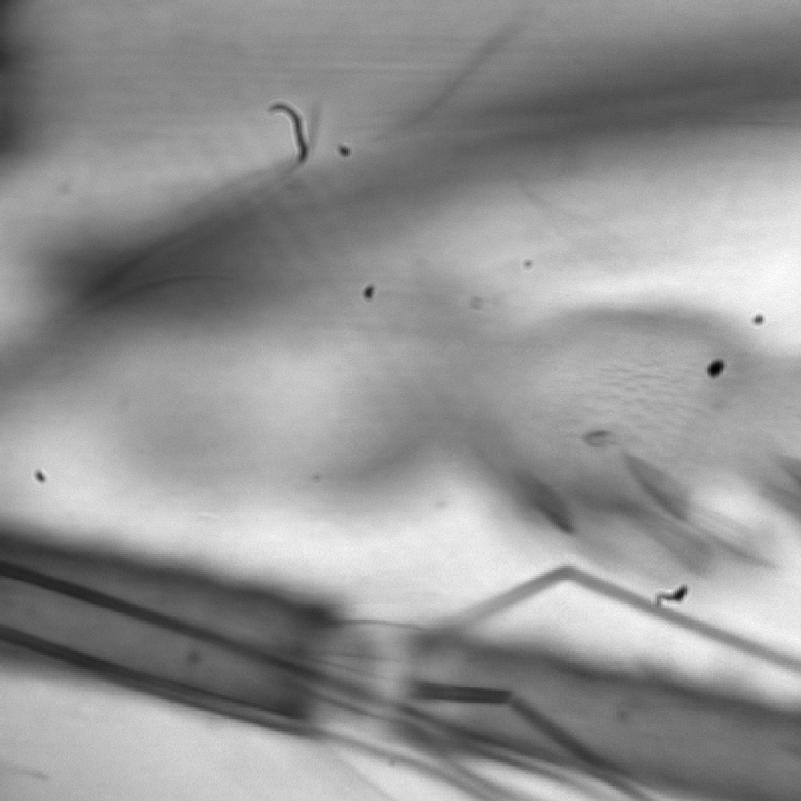}
	\caption{}
	\end{subfigure}
	\caption{Reconstructed parallax view images with LFMI using 2 and 7 images respectively~(see \href{Visualization5.avi}{Visualization 5} and \href{Visualization7.avi}{Visualization 7}), and with LFBP using 2 and 7 images~(see \href{Visualization6.avi}{Visualization 6} and \href{Visualization8.avi}{Visualization 8}).}
	\label{fig:LF_diffNum}
\end{figure}




Due to the convenience of capturing multiple focal plane sweeping images with the proposed system, we also show the comparison with LFMI and LFBP using different number of images. The results are shown in Fig.~\ref{fig:LF_diffNum}. Fig.~\ref{fig:LF_diffNum}(a)(b) show the parallax view images with LFMI using 2 and 7 captured images respectively. Fig.~\ref{fig:LF_diffNum}(c)(d) show the reconstructed images using LFBP. More detail can be observed from the videos of  visualization 5, visualization 6, visualization 7, and visualization 8. 
In LFMI, the light field moment can more accurate as the increasing number of the used images. However,  the approximate Gaussian function makes it difficult to get details of the light field. Therefore, the light field reconstructed using more images isn't distinctly improved compared to using 2 images. As shown in Fig.~\ref{fig:LF_diffNum}(a)(b).
The LFBP reconstruction can be considered as a averaging filter, which increasing the weight of the light in reconstruction direction. This filter might be simple and make the reconstructed images not distinct enough because of the crosstalk from the other points. Therefore, the quality of the reconstruction depends much on the number of the used images, as shown in Fig.~\ref{fig:LF_diffNum}(c)(d). This results are more persuasion, because in the capturing process, there is no other factors that affect the quality of the captured images. 


\section{Conclusion}
We have proposed a focal plane sweeping capture system with defocus modulation using a SLM. With this system, the time cost for capturing a large amount of focal plane sweeping images is efficiently reduced. And the accuracy of the captured images is increased because there is no mechanical movement during the capture process. The captured images were used to perform light field reconstruction with two techniques, i.e., LFMI and LFBP. Because of the controllability of the system PSF, it is easier to capture images that meet the special requirements of either LFMI or LFBP. 

It should be mentioned that the PSF of the imaging system can also be other distribution functions rather than Gaussian. In this case, the Gaussian distribution function in the LFMI equation should be modified to the corresponding PSF function. The imaging system in our paper is a microscopic, this can also be extended to conventional digital camera system. In that case, the SLM can be replaced by an electrically tunable lens for colorful imaging. The SLM in the proposed system in this paper can also be replaced by an electrically tunable lens for color imaging.

\section{Acknowledgments}
This work was supported by National Natural Science Foundation of China~(NSFC) (61327902, 61377005), Chinese Academy of Sciences~(CAS) (QYZDB-SSW-JSC002), and Natural Science Foundation of Shanghai~(NSFS) (17ZR1433800).



\end{document}